\documentclass[onecolumn,preprintnumbers,amsmath,amssymb]{revtex4}

\usepackage{graphicx}
\usepackage{dcolumn}
\usepackage{bm}

\begin{document}
\title{Theory of Electronic relaxation in solution: Exact solution in case of parabolic potential with a sink of finite width}
\author{Swati Mudra*}
\affiliation{School of Basic Sciences, Indian Institute of Technology Mandi, Kamand, Himachal Pradesh 175005, India.}
\author{Hemani Chhabra}
\affiliation{Theoretical Chemistry Group, University of Oxford, Oxford, UK.}
\author{Aniruddha Chakraborty} 
\affiliation{School of Basic Sciences, Indian Institute of Technology Mandi, Kamand, Himachal Pradesh, 175005, India.}
\date{\today }
\begin{abstract} 
\noindent We give a general method for finding the exact solution for the problem of electronic relaxation in solution, modeled by a particle undergoing diffusive motion under a potential in the presence of a sink of finite width. The solution requires the knowledge of the Green's function in Laplace domain in the absence of any sink. We find the exact solution for the case of parabolic potential. This model has considerable improvement over the existing models for understanding non-radiative electonic relaxation of a molecule in solution, in fact this is the first model where a simple analytical solution is possible in the case of a sink of finite width. 
\end{abstract}

\maketitle

\section{Introduction}
\noindent Relaxation of a polar molecular in a solvent is a matter of interest which seeks attention of both experimentalist and theoreticians \cite{Ani1,Kls1,Bagchi,SK,Ani2}. There are different cases possible when a molecule absorbs some light in a solution \cite{Robin}. The molecule will go in higher potential energy surface  and perform random motion due to the effect of solvent \cite{Ani1, Kls1, Ani2}. The relaxation may occur in two ways, either radiative from anywhere of the potential energy surface or nonradiative from a certain region of that potential energy surface \cite{Robin}. Smoluchowski equation is widely used to calculate the Survival probability of the molecule in such cases \cite{Amb}. Analytical solution of Smoluchowski equation is known only for a few potentials with a delta function sink \cite{Kls1,Kls2,Kls3}. Here we are giving an exact solution for survival probability using Smoluchowski equation for a parabolic potential with a sink of finite width. The same expression for survival probability is then used to derive analytical expressions for rate constants.

\section{Smoluchowski Equation for a parabolic potential with a sink}

\noindent  We are using Smoluchowski equation to solve our model which describes the time evolution of the number density of the particles with distance x and time t. Here P(x,t) is the number density of the particles in the excited state potential energy level or we can say the survival probability of particles in excited state. $k_0$  is the non radiative decay rate and $k_r$ is the radiative decay rate of the particle. $S(x)$ is the localized sink which is non zero between $x_c - \epsilon$ to $x_c+\epsilon$ . V(x) is the potential in which the particle is executing its motion. The Smoluchowsky equation for our model is
\begin{equation}
\frac{\partial P(x,t)}{\partial t} = (L - k_{r} - k_{o} S(x)) P(x,t).
\end{equation}
where
\begin{equation}
    L = D\frac{\partial^2}{\partial x^2} + \frac{D}{k_{b}T}\frac{\partial}{\partial x}\left(\frac{\partial V(x)}{\partial x}\right).
\end{equation}
Here the potential we are taking is parabolic so
\begin{equation}
    V(x) = \frac{Bx^2}{2}.
\end{equation}
by substituting all the terms we get
\begin{equation}
\frac{\partial P(x,t)}{\partial t} = D\frac{\partial^2 P(x,t)}{\partial x^2} + \frac{DB}{k_{b}T} \frac{\partial}{\partial x} x P(x,t) - k_{o} S(x) P(x,t) - k_{r} P(x,t).
\end{equation}

\section{Exact Solution of the equation}
\noindent The Laplace transform of $P(x,t)$ is given by 
\begin{equation}
\tilde P(x,s)= \int^\infty_0 P(x,t) e^{-st} dt.
\end{equation}
so the Laplace transformation of Eq.(4) is given by\\
\begin{equation}
\left[s {\tilde P}(x,s)-D\frac{\partial^2{\tilde P}(x,s)}{\partial x^2} - \frac{DB}{k_{b}T}\frac{\partial}{\partial x} \left(x {\tilde P}(x,s)\right) + k_{o} S(x){\tilde P}(x,s)+{k_{r}}{\tilde P}(x,s)\right] =  P(x,0).
\end{equation}
In the following we assume that $S(x)$ is non-zero for a short range of $x$ - values around $x = x_c$, so we can replace $S(x)$ by $S(x_c)$ in the above equation ignoring all the higher order terms in the Taylor series expansion of $S(x)$. As a result we can replace the term $S(x) {\tilde P}(x,s)$ by $S(x_c) {\tilde P}(x_c,s) f(x)$, where $f(x)$ equals to $1$ for $x$ values between $x_c - \epsilon$ to $x_c+\epsilon$ and $f(x)$ equals to zero otherwise, where $\epsilon$ is a very small positive number. So Eq. (6) becomes
\begin{equation}
\left[s {\tilde P}(x,s)- D\frac{\partial^2{\tilde P}(x,s)}{\partial x^2} - \frac{DB}{k_{b}T}\frac{\partial}{\partial x}\left( x {\tilde P}(x,s)\right)+ k_{0} S(x_c){\tilde P}(x_c,s) f(x)+k_{r}{\tilde P}(x,s)\right] =  P(x,0).
\end{equation}
The solution of this equation in terms of Green's function $G(x,s|x_0)$ is given below
\begin{equation}
\tilde P(x,s)= \int^\infty_{-\infty} dx_{0}G(x,s+k_{r}|x_0)P(x_0,0) - k_{0} S(x_c){\tilde P}(x_c,s)\int^{\infty}_{-\infty} dx_{0}G(x,s+k_{r}|x_0)f(x_0).
\end{equation}
The above equation can further be simplified by using the properties of $f(x)$ as given below
\begin{equation}
\tilde P(x,s)= \int^\infty_{-\infty} dx_{0} G(x,s+k_{r}|x_0)P(x_0,0) - k_{0} S(x_c){\tilde P}(x_c,s)\int^{x_c+\epsilon}_{x_c -\epsilon} dx_{0}G(x,s+k_{s}|x_0).
\end{equation}
The above equation can further be simplified to
\begin{equation}
\tilde P(x,s)= \int^\infty_{-\infty} dx_{0}G(x,s+k_r|x_0)P(x_0,0) - 2 k_{0} \epsilon S(x_c){\tilde P}(x_c,s)G(x,s+k_r|x_c).
\end{equation}
Now put $x=x_c$ in equation (10), we get
\begin{equation}
\tilde P(x_c,s)= \int^\infty_{-\infty} dx_{0}G(x_c,s+k_r|x_0)P(x_0,0) - 2 k_{0} \epsilon S(x_c){\tilde P}(x_c,s)G(x_c,s+k_r|x_c).
\end{equation}
Now we solve the above equation for $\tilde P(x_c,s)$ to get
\begin{equation}
\tilde P(x_c,s)= \frac{\int^\infty_{-\infty} dx_{0}G(x_c,s+k_r|x_0)P(x_0,0)}{1+2 k_{0} \epsilon S(x_c)G(x_c,s+k_r|x_c)}.
\end{equation}
Now substitute it back into Eq. (10), we get
\begin{equation}
\tilde P(x,s)= \int^\infty_{-\infty} dx_{0} \left(G(x,s+k_r|x_0)- 2 k_{0} \epsilon S(x_c) G(x,s+k_r|x_c)G(x_c,s+k_r|x_0)[1+ 2 k_{0} \epsilon S(x_c) G(x_c,s+k_r|x_c)]^{-1}\right) P(x_0,0).
\end{equation}
In the following we will assume $S(x_c)=1$ so that we get
\begin{equation}
\tilde P(x,s)= \int^\infty_{-\infty} dx_{0} \left(G(x,s+k_r|x_0)- 2 k_{0} \epsilon G(x,s+k_r|x_c)G(x_c,s+k_r|x_0)[1+ 2 k_{0} \epsilon G(x_c,s+k_r|x_c)]^{-1}\right) P(x_0,0).
\end{equation}
Now we calculate the survival probability $P_e(t) =\int^\infty_{-\infty} dx P(x,t)$, but it is very difficult so Instead that we can easily calculate the Laplace transform of $ P_e(t)$ directly. Which is $P_e(s)$ associated to $P(x,s)$ by 
\begin{equation}
P_e(s) = \int^\infty_{-\infty} dx {\tilde P}(x,s).
\end{equation} 
Using Eq. (14) and the fact that $\int^\infty_{-\infty} dx_{0} (G(x,s|x_0) = 1/s$, we get
\begin{equation}
P_e(s)=\frac{1}{s+k_r}\left[1-[1+2 k_{0} \epsilon G(x_c,s+ k_r|x_c)]^{-1} 2 k_{0} \epsilon\times \int^\infty_{-\infty} dx_0 G (x_c,s+k_r|x_0)P(x_0,0)\right].
\end{equation}
The average and long time rate constants can be easily derived from the formula of $P_e(s).$ Thus, $k^{-1}_I =P_e(0)$ and $k_L$ = negative of the pole of $P_e(s),$ which is close to the origin. From Eq. (16), we obtain
\begin{equation}
k^{-1}_I =\frac{1}{k_{r}}\left[1- [1+2 k_{0} \epsilon G(x_c,k_r|x_c)]^{-1} 2 k_{0} \epsilon\times \int^\infty_{-\infty} dx_0 G(x_c,k_r|x_0)P(x_0,0)\right].
\end{equation}
Thus $k_I$ depends on the initial probability distribution $P(x,0)$ whereas $k_L = - $ pole of $\left([ 1+2 k_0 \epsilon\; G(x_c,s+k_r|x_c)][s+k_r]\right)^{-1}$, the one which is closest to the origin, on the negative $s$ - axis, and is independent of the initial distribution $P(x_0,0)$.
The $G_0(x,s;x_0)$ can be found out by using the following equation \cite{Kls2}:
\begin{equation}
\left(s - {\cal L}\right) G_{0}(x,s;x_0)= \delta (x - x_0). 
\end{equation}
Using standard method \cite{Hilbert} to obtain.
\begin{equation}
G_0(x,s;x_0)=F(z,s;z_0)/(s+k_r)
\end{equation}
with
\begin{equation}
F(z,s;z_0)= D_\nu(-z_<)D_\nu(z_>)e^{(z_0^2-z^2)/4}\Gamma(1-\nu){[B/(2 \pi D)]}^{1/2}.
\end{equation}
In the above, $z$ defined by $z = x (D/B)^{1/2}$  and $z_j = x_j (D/B)^{1/2}$, $\nu  = —s B$ and $\Gamma(\nu)$ is the gamma function. Also, $z_{<}= min(z, z_0)$ and $z_{>}= max(z, z_0)$. $D_{\nu}$ represent parabolic cylinder functions. To get an understanding of the behavior of $k_I$ and $k_L$, we assume the initial distribution $P^0_e(x_0)$ is represented by $\delta(x-x_0)$. Then, we get 
\begin{equation}
{k_I}^{-1}= (k_r)^{-1}\left(1 - \frac{2 k_{0} \epsilon F(z_s,k_r|z_0)}{k_r+ {2 k_{0} \epsilon}F(z_s,k_r|z_s)} \right).
\end{equation}
Again
\begin{equation}
k_L= k_r - [ values \; of \; s \; for \; which \;\; s+ {2 k_{0} \epsilon} F(z_s,s|z_s)=0].
\end{equation}
We should mention that $k_I$ is dependent on the initial position $x_0$ and $k_r$ whereas $k_L$ is independent of the initial position.
In the following , we assume $k_r \rightarrow $ 0, in this limit we arrive at conclusions, which we expect to be valid even when $k_r$ is finite. Using the properties of $D_v{(z)}$, we find that when $k_r \rightarrow 0, F{(z_s,k_r|z_0)}$ and $F{(z_s,k_r|z_s)}\rightarrow exp(-z_s^2/2)/[B/(2 \pi D)]^{1/2}$so that
\begin{equation}
2 k_0 \epsilon F{(z_s,k_r|z_0)}/[k_r + 2 k_0 \epsilon F{(z_s,k_r|z_s)}]\rightarrow 1.
\end{equation}
\\Hence 
\begin{equation}
k_I^{-1}=-{[\frac{\partial}{\partial k_r}\left[\frac{2 k_0 \epsilon F(z_s,k_r|z_0)}{k_r + 2 k_0 \epsilon F(z_s,k_r|z_s)}\right]}_{k_r \rightarrow 0}.
\end{equation} 
If we take $z_0 < z_s $, so that the particle is initially placed to the left of sink. Then \\
\begin{equation}
k_I^{-1}= e^{{z_s}^2/2}/\{2 k_0 \epsilon {[B/(2 \pi D)]}^{1/2}\}+ \left[\frac{\partial}{\partial k_r}\left[\frac{e^{[(z_0^2-z_s^2)/4]}D_v{(-z_0)}}{D_v{(-z_s)}}\right]\right]_{v=0}.
\end{equation} 
After simplification
\begin{equation}
k_I^{-1}= e^{{z_s}^2/2}/\{2 k_0  \epsilon {[B/(2 \pi D)]}^{1/2}\}+ \left(\int_{z_0}^{z_s} dz e^{(z^2/2)}\left[1+erf(z/\sqrt{2}\right]\right)(\pi/2)^{1/2}/B.
\end {equation}
The long-term rate constant $k_L$ is determined by the value of $s$, which satisfy $s+k_0 F(z_s,s|z_s)=0$. This equation can be written as an equation for $\nu (= -s/B)$
\begin{equation}
\nu = D_\nu(-z_c)D_\nu(z_c)\Gamma(1-\nu)2 k_0 \epsilon {[B/(2 \pi D)]}^{1/2}.
\end{equation}
For integer values of $\nu$, $D_\nu(z)=2^{-\nu/2}e^{-z^2/4}H_{\nu}(z/\sqrt{2})$, $H_{\nu}$ are Hermite polynomials. $\Gamma(1-\nu)$ has poles at $\nu = 1,2, . . . .$. Our interest is in $\nu \in [0, 1]$, as $k_L = B \nu$ for $k_r =0$. If $ 2 k_0 \epsilon {[B/(2 \pi D)]}^{1/2}\ll 1$, or $z_c \gg 1$ then $\nu \ll 1$ and one can arrive
\begin{equation}
\nu = D_0(-z_s)D_0(z_s)2 k_0 \epsilon {[B/(2 \pi D)]}^{1/2}.
\end{equation}
and hence 
\begin{equation}
k_L = e^{{{-z_s}^2}/2} 2 k_0 \epsilon {[B/(2 \pi D)]}^{1/2}.
\end{equation}
In this limit, the rate constant $k_L$ exhibits Arrhenius type activation.

\section{Conclusions:}

\noindent In this paper, we proposed a method to solve the electronic relaxation problem in a parabolic potential for a finite localized sink. This method is purely analytical and it gives a general solution for survival probability of molecule in a parabolic potential energy surface in presence of a sink of finite width. Both the rate constants varies with the width of the sink. Long time rate constant is directly proportional to the width of the sink. The long time rate constant is following the Arrhenius equation. We conclude $K_{L}$ is directly proportional to the force constant B and inversely proportional to the diffusion constant D.  

\section{Acknowledgments:}
\noindent One of the author (S.M.) would like to thank IIT Mandi for HTRA fellowship and the other author thanks IIT Mandi for providing PDA grant.

\end{document}